\documentstyle[11pt,aaspp]{article}

\tighten

\slugcomment{\hfill COBE Preprint \# 96-03, Submitted to {\it ApJ Letters}}

\begin{document}

%%%%%%%%%%%%%%%%%%%%%%%%%%%%%%%%%%%%%%%%%%%%%%%%%%%%%%%%%%%%%

\def\pmb#1{\setbox0=\hbox{#1}%
  \kern0.00em\copy0\kern-\wd0
  \kern0.03em\copy0\kern-\wd0
  \kern0.00em\raise.04em\copy0\kern-\wd0
  \kern0.03em\raise.04em\copy0\kern-\wd0\box0 }

\def\pp{\parshape=2 -0.25truein 6.75truein 0.5truein 6truein}

\def\ref #1;#2;#3;#4;#5{\par\pp #1 #2, #3, #4, #5}
\def\book #1;#2;#3{\par\pp #1 #2, #3}
\def\rep  #1;#2;#3{\par\pp #1 #2, #3}

\def\undertext#1{$\underline{\smash{\hbox{#1}}}$}
\def\simlt{\lower.5ex\hbox{$\; \buildrel < \over \sim \;$}}
\def\simgt{\lower.5ex\hbox{$\; \buildrel > \over \sim \;$}}

\def\bCCMB{{\bf C}_{CMB}}
\def\bC{{\bf C}}
\def\bp{{\pmb {$\psi$}}}
\def\bG{{\pmb {$\Gamma$}}}
\def\by{{\bf y}}
\def\bof{{\bf f}}
\def\ba{{\bf a}}
\def\bc{{\bf c}}
\def\bI{{\bf I}}
\def\bL{{\bf L}}
\def\bW{{\bf W}}
\def\bS{{\bf S}}
\def\bN{{\bf N}}
\def\bC{{\bf C}}

\def\etal{{\it et~al.\ }}
\def\np{\vfill\eject}
\def\cl{\centerline}
\def\u{\undertext}
\def\ul{\underline}
\def\noi{\noindent}
\def\bs{\bigskip}
\def\ms{\medskip}
\def\ss{\smallskip}
\def\i{\item}
\def\ii{\item\item}
\def\ob{\obeylines}
\def\l{\line}
\def\hrf{\hrulefill}
\def\hf{\hfil}
\def\q{\quad}
\def\qq{\qquad}
\newcommand{\uk}{$\mu$K}

%%%%%%%%%%%%%%%%%%%%%%%%%%%%%%%%%%%%%%%%%%%%%%%%%%%%%%%%
%%%%%%%%%%%%%%%%%%%%%%%%%%%%%%%%%%%%%%%%%%%%%%%%%%%%%%%%

\title{Power Spectrum of Primordial Inhomogeneity Determined
from the 4-Year {\it COBE\thinspace}\altaffilmark{1} DMR Sky Maps}

\author{
K.M. G\'orski\altaffilmark{2,3,4}, 
A.J. Banday\altaffilmark{2,5},
C.L. Bennett\altaffilmark{6},
G. Hinshaw\altaffilmark{2},
A. Kogut\altaffilmark{2},
G.F. Smoot\altaffilmark{7}, 
and~E.L.~Wright\altaffilmark{8} 
}

\noindent
\altaffiltext{1}{The National Aeronautics and Space Administration/Goddard 
Space Flight Center (NASA/GSFC) is responsible for the design, development, 
and operation of the {\it Cosmic Background Explorer (COBE)}. Scientific 
guidance is provided by the {\it COBE} Science Working Group. GSFC is also 
responsible for the analysis software and for the 
production of the mission data sets.}
\altaffiltext{2}{Hughes/STX Corporation, 
LASP, Code 685, NASA/GSFC, Greenbelt, Maryland 20771.}
%\altaffiltext{2}{Hughes/STX Corporation, 
%Laboratory for Astronomy and Solar Physics, 
%Code 685, NASA/Goddard Space Flight Center, Greenbelt, Maryland 20771.}
\altaffiltext{3}{on leave from Warsaw University Observatory, 
                 Aleje Ujazdowskie 4, 00-478 Warszawa, Poland.}
\altaffiltext{4}{e-mail: {\it gorski@stars.gsfc.nasa.gov}}
\altaffiltext{5}{Max Plank Institut fur Astrophysik, 85740 Garching Bei
Munchen, Germany.}
\altaffiltext{6}{Laboratory for Astronomy and Solar Physics, 
Code 685, NASA/GSFC, Greenbelt, Maryland 20771.}
\altaffiltext{7}{LBL, SSL, \& CfPA, Bldg 50-351, University of California, 
                 Berkeley CA 94720.}
\altaffiltext{8}{UCLA Astronomy Department, Los Angeles CA 90024-1562.}
%;\altaffiltext{9}{UCSB Physics Department, Santa Barbara CA 93106.}

\begin{abstract}
Fourier analysis and power spectrum estimation of the 
cosmic microwave background anisotropy on an incompletely
sampled sky
developed by G\'orski (1994)
has been
applied to the high-latitude portion of the 
4-year {\it COBE} DMR 31.5, 53 and 90 GHz sky maps. 
Likelihood analysis using 
newly constructed Galaxy cuts (extended beyond $\vert b\vert = 20\ $deg to 
excise the known foreground emission) 
and simultaneously correcting for the faint high latitude galactic foreground
emission
is conducted 
on the DMR  sky maps 
pixelized in both
ecliptic and galactic coordinates.
The 
Bayesian  power spectrum estimation from the foreground corrected 4-year
{\it COBE} DMR data renders 
$n \sim 1.2 \pm 0.3$, and 
$Q_{rms-PS} \sim 15.3^{+3.7}_{-2.8}\ \mu$K (projections of the two-parameter
likelihood).
These results 
are consistent with the Harrison-Zel'dovich $n=1$ model 
of amplitude $Q_{rms-PS} \sim 18\ \mu$K
detected with significance  exceeding $14\sigma$ ($\delta Q /Q \simlt 0.07$).
(A small power spectrum amplitude drop below the published 2-year 
results
is predominantly due to the application of the new, extended Galaxy cuts.)

\end{abstract}


\section{INTRODUCTION}

Following the seminal {\it COBE} DMR discovery of the anisotropy of
the cosmic microwave
background (CMB) 
after one year (1989/90) of observations 
(Smoot \etal 1992, Bennett \etal 1992, Wright \etal 1992, Kogut \etal 1992)
the instrument continued operating until late 1993. 
Intermediate, 2-year results
of the mission were reported by Bennett \etal (1994) (also see e.g.
G\'orski \etal 1994, Wright \etal 1994).
An overview of the 4-year results is provided by Bennett \etal (1996).
In this paper we apply the CMB anisotropy power spectrum estimation technique
introduced by G\'orski (1994) to the definitive, 4-year 
31.5, 53, and 90 GHz sky maps resulting from 
the {\it COBE} DMR mission.

The {\it COBE} DMR-detected anisotropy of the CMB 
provides a unique opportunity to measure the spatial distribution
of the inhomogeneities in the universe on the comoving scales ranging from
a few hundred Mpc up to 
the present horizon size
(inaccessible to any other astronomical observations)
during the embryonic stage of their evolution, thereby avoiding
the 
complications of 
(cosmologically) recent
evolution of most astrophysical systems.
There are, however, important experimental/observational
limitations on the extent to which cosmologically interesting quantities
can be extracted from the DMR data.
These are imposed by our understanding
of systematic effects, instrumental noise properties, and 
non-cosmological foreground signals peculiar to our location
in the Galaxy and
in the universe, and were investigated as follows.
Kogut \etal (1996c) have addressed systematic effects
in considerable detail and found them to be unimportant relative to the 
accuracy to which cosmological parameters can be determined from the DMR 
4-year data.
G\'orski \etal (1996) provide a comprehensive description 
of the noise properties of the DMR sky maps
and demonstrate the 
sufficiency of the DMR noise models
%Gaussian model with small 60$^{\circ}$-ring correlations 
(see also Lineweaver \etal 1994).
Emission from the galactic plane 
cannot yet be modelled to adequate precision
and must be excised from the data (see \S2 and Banday \etal 1996a).
Foreground galactic emission at 
high latitude (Bennett \etal 1992,
Kogut \etal 1996a,b) 
can be  partially 
accounted for by using Galaxy-dominated spatial templates at 
frequencies where the CMB does not dominate (see \S3, and Kogut \etal 1996a,b). 
The potential contamination of the CMB anisotropy by the astrophysical
foregrounds outside of our Galaxy is demonstrated to be negligible 
in Banday \etal (1996b).

In this work we focus on a general characterization of
the 4-year DMR sky maps in terms of a power spectrum
of the angular distribution of the CMB temperature fluctuations
and attempt to estimate
its sensitivity to a plausible 
foreground emission contamination. We follow and further develop 
the method introduced by G\'orski (1994, hereafter G94a) and 
applied to the 2-year DMR data in G\'orski \etal (1994, hereafter G94b).
As a useful power spectrum parametrization, 
we utilize a power-law family
specified by $Q_{rms-PS}$ and $n$ --- the amplitude and 
shape parameters 
%(G94a, 
(Bond \& Efstathiou 1987, Fabbri, Lucchin,
\& Matarrese 1987), but we
abbreviate  the symbol $Q_{rms-PS}$ and use $Q$ instead.
The results of the analysis involving the CMB spectra
of specific large scale structure models will be presented elsewhere.
Hereafter, bold letters denote matrices (upper case) and vectors
(lower case), and $p$ is a pixel label.

\section{DATA}

Our published analysis of the 2-year DMR data 
%using Fourier decomposition
%on the cut sky 
(G94b)
involved a simple
$\vert b\vert > 20^\circ$ Galaxy cut, which retained 
the relatively bright foreground emission
from the Scorpius-Ophiucus and Taurus-Orion regions (de Vaucouleurs 1955).
In this work we employ new, extended galactic 
cuts (see Banday  \etal 1996a) which further excise these regions
from the data used for the inference of the cosmological CMB anisotropy.
We consider both extended 
$\vert b\vert > 20^\circ$ (denoted `$20+$') and
extended 
$\vert b\vert > 30^\circ$ (`$30+$') Galaxy cuts.
%in order to test the stability of the derived power spectrum
%parameters with respect to data selection on the sky,  
We analyze the DMR sky maps 
in both ecliptic (`E') and galactic (`G') pixelization.
To facilitate a comparison of  the 4-year results with the published 
analyses of the 2-year data we re-apply
the simple $\vert b\vert > 20^\circ$ Galaxy cut.
From a total of 6144 data pixels on the entire sky,
the following numbers of
pixels remain for analysis after applying the different galactic cuts:
4038 for E20,
4016 for G20,
3890 for E20$+$, 
3881 for G20$+$, 
3039 for E30$+$, and 
3056 for G30$+$.  

%The 31.5, 53 and 90 GHz four year cut-sky DMR maps 
%are jointly analyzed in this work.  
We form weighted average maps at each frequency 
by combining the 
%individual 
A and B channels,
% as follows
%(in thermodynamic units)
$$\matrix{
\Delta_{31} = 0.611\,\Delta_{31A} + 0.389\,\Delta_{31B},\cr
\Delta_{53} = 0.579\,\Delta_{53A} + 0.421\,\Delta_{53B},\cr
\Delta_{90} = 0.382\,\Delta_{90A} + 0.618\,\Delta_{90B},\cr }
\eqno(1)$$
%where all temperatures are in thermodynamic units.  
%These weights 
%were chosen to 
minimizing the cut-sky-averaged 
noise variance per pixel in the combined maps. 
The weighted 31.5, 53, and 90 GHz 4-year cut-sky DMR maps 
are analyzed jointly (equivalent to coaddition)
with and without the foreground correction, or
linearly combined (after foreground correction, indicated by tilde) 
with the weights designed
to remove any leftover free-free emission:
$$\widetilde{\Delta}_{CMB} = -0.302\,
\widetilde{\Delta}_{31} + 0.633\,
\widetilde{\Delta}_{53} + 0.669\,
\widetilde{\Delta}_{90}.
\eqno(2)$$
%where tilde denotes the foreground correction  described below (\$3).
The weights given to both 31.5 and 90 GHz maps 
in this linear combination technique
are considerably
larger than 
%the weights used for these maps 
those in 
the coadded technique
($\sim 0.08$ and $\sim 0.26$, respectively). 
As a result, the noise is larger in the
linear combination map $\widetilde{\Delta}_{CMB}$,
slightly exceeding the rms noise in the 90 GHz map alone. 

Fourier analysis of the cut-sky maps 
is performed (see G94a for details) in the 961-dimensional linear space
spanned by the orthonormal functions $\bp$, which are linear combinations
of spherical harmonics with $\ell \le 30$.
%(G\'orski 1994).
A sky map at frequency $\nu$, $\Delta_\nu(p)$ (where $p$ is the pixel label)
is Fourier decomposed as follows
%in the $\bp$ 
%basis are given by 
$$
c_{i,\nu} = {4\pi\over 6144} \sum_{p\in \{cut\; sky\}} \Delta_\nu (p)\;
\psi_i(p) 
\;\;\;\; {\rm  and}\;\;\;
\Delta_\nu (p)\vert_{p\in{\{cut\;sky\}}}
= \bc_{\nu} ^T \cdot \bp (p). 
%\bc_\nu = \langle \Delta_\nu \;\bp\rangle_{\{cut\; sky\}},
%\approx \int_{\{cut\; sky\}} d\Omega \Delta_\nu (\Omega)\;
%\psi_i(\Omega),
\eqno(3)
$$
Fourier coefficients $c_i$ are linear in pixel temperatures, 
and account for the cut-sky mode coupling explicitly. 
Figure 1 (Color plate) displays an example of the $\bp$ basis function
constructed for various versions of the Galaxy cut used in this work.

The dominant component (in rms per pixel) of the raw,
dipole-subtracted, cut-sky DMR maps is the instrument generated noise.
A correct statistical modeling of this contaminant is
essential for any method attempting to quantify the cosmological CMB
anisotropy. A detailed analysis of noise properties of the entire
DMR data set is discussed in G\'orski \etal (1996). Here we employ the 
noise modeling in Fourier space as described in G94a,b.
The noise content of the 4-year DMR data is characterized
by the rms noise amplitudes per Fourier mode of approximately
10.5, 3.4, and 5.5 $\mu$K at
31.5, 53, and 90 GHz, respectively.
The coaddition of  all available measurements at three frequencies renders
a map with the rms noise amplitude per Fourier mode of $\sim 2.7\ \mu$K.

\section{LIKELIHOOD ANALYSIS OF FOREGROUND CORRECTED DMR SKY MAPS}

Following G94a,b, we conduct Bayesian inference from the {\it COBE} DMR data
of the power spectrum of primordial CMB anisotropy 
%(as described in G94a,b)
under the usual assumption that the measurement 
(here the joint vector of coefficients of Fourier
decomposition of the 31.5, 53, and 90 GHz sky maps,
$\widehat\bc^T_{\uplus} \equiv (\widehat\bc^T_{31},\widehat\bc^T_{53},
\widehat\bc^T_{90})$) 
is a sum of a signal --- primordial CMB 
anisotropy, and noise --- a random, receiver generated contamination,
and that both these components are Gaussian stochastic variables.
Hence,
the likelihood function is 
{\it exactly} represented
as a multivariate Gaussian in variable $\bc_{\uplus}$.
The likelihood density is fully specified by the 
Fourier space
correlation matrices 
of signal, $\bC_{S}$,  
a function of the assumed (and tested)
power spectrum of the CMB
anisotropy,
and noise (at a given frequency $\nu$), 
$\bC_{N_{\nu}}$,
which are are computed as described
in G94a,b. 
No approximations or Monte Carlo simulations are required for evaluation of 
the likelihood function, and this power spectrum parameter estimation 
technique, bi-linear in the measurement values of temperature perturbations,
is statistically unbiased.

This formulation of the likelihood problem  
allows one naturally to address a problem of contamination of the
CMB anisotropy data by the diffuse foreground emission. 
Assume (generally) that the anisotropy, $\bc_{\nu}$, 
is measured at $M$ frequencies $\nu$, and 
the foreground emission can be measured or modeled as a certain
number, $N$, of non-stochastic 
spatial templates $F_k(p)$ (whose Fourier transforms
are $\bof_k$, $k=1,\ldots,N$).
The template corrected joint data vector,
$\widetilde{\bc}_{\uplus}$, 
and its covariance matrix,
$\bC_{SN\uplus}$,
are
$$\widetilde{\bc}_{\uplus}          =
\left( \matrix{ 
\widehat{\bc}_{\nu_1} - \sum_{k=1}^{k=N} \alpha_{\nu_1}^k {\bof_k} \cr
\cdots                                                               \cr
\widehat{\bc}_{\nu_{M}} -\sum_{k=1}^{k=N} 
\alpha_{\nu_{M}}^k {\bof_k}                                      \cr} 
\right), \;\;
\bC_{SN\uplus} =
\left\langle \widetilde{\bc}_{\uplus}\! \cdot 
              \widetilde{\bc}_{\uplus}^{\, T} \right\rangle =
%\left( \matrix{ \bC _{SN31} \hfill & \bC _{S}    \hfill & \bC _{S}   \hfill \cr
%                \bC _{S}    \hfill & \bC _{SN53} \hfill & \bC _{S}   \hfill \cr
%                \bC _{S}    \hfill & \bC _{S}    \hfill & \bC _{SN90}\hfill\cr }
\left( \matrix{ \bC _{SN_{\nu_1}} \hfill& \cdots\hfill & \bC _{S}  \hfill \cr
                \cdots    \hfill        & \cdots\hfill & \cdots    \hfill \cr
                \bC _{S}  \hfill        & \cdots\hfill & \bC _{SN_{\nu_{M}}}\hfill\cr }
\right), 
\eqno(4)
$$
where the matrix $\bC_{SN_{\nu_i}} = \bC_S + \bC_{N_{\nu_i}}$
specifies the Gaussian probability distribution of the 
receiver noise contaminated 
theoretical CMB anisotropy signal at a given frequency $\nu_i$.
With these definitions the likelihood function takes the 
usual form  
$$
P(\widetilde{\bc}_{\uplus}) \propto
\exp \Bigl( -\, \widetilde{\bc}_{\uplus}^{\, T} \cdot 
\bC_{SN\uplus}^{-1} \cdot 
\widetilde{\bc}_{\uplus}\, \big/\ 2\, \Bigr)\bigg/
\sqrt{ {\rm det}\; \bC_{SN\uplus}}.
\eqno(5)
$$
This is a function of all parameters used to describe the CMB power spectrum 
and $M \times N$ coefficients $\alpha_\nu ^i$ used 
in the foreground correction. 
For a given CMB power spectrum 
(i.e. a fixed $\bC_S$) the likelihood can be 
maximized with respect to parameters $\alpha_\nu^i$ as a usual 
$\chi^2$ fitting problem (naturally, 
the dimensionality of the data vector is presumed to be large compared to both
$M$ and $N$).
From the maximum likelihood equations, 
$\partial\ {\rm ln} P / \partial \alpha_\nu ^i = 0$, one derives the following
set of linear equations (for a given  $\nu$  and $k$):
$$\sum_{j, \nu'}\ {\bof}^T_k \cdot \left( \bC_{SN\uplus}^{-1} \right)_{\nu \nu'}
\cdot
{\bof}_j \; \alpha^j_{\nu'} = 
\sum_{\nu''}\ {\bof}^T_k \cdot \left( \bC_{SN\uplus}^{-1} \right)_{\nu \nu''}
\cdot
{\widehat \bc}_{\nu''}, 
\eqno(6)$$
which render the maximum likelihood solutions $\widehat{\alpha}_\nu^k$.
The parameter 
%uncertainty 
covariance matrix (dimension $(M\times N)^2$) is 
$$ M^{-1}_{\alpha \alpha'} \equiv
%\left\langle 
%\left({\alpha}_\nu^i - \widehat{\alpha}_\nu^k\right)
%\left({\alpha}_{\nu'}^{i'} - \widehat{\alpha}_{\nu'}^{k'}\right)
%   \right\rangle^{-1} =
-{\partial ^2 {\rm ln} P\over \partial \alpha_\nu^k \partial 
\alpha_{\nu'}^{k'}}=
{\bof}^T_k \cdot \left( \bC_{SN\uplus}^{-1} \right)_{\nu \nu'}
\cdot
{\bof}_{k'}. 
\eqno(7)$$
In both equations  $\left( \bC_{SN\uplus}^{-1} \right)_{\nu \nu'}$
is a square portion 
%segment 
of the inverse covariance matrix  $\bC_{SN\uplus}^{-1}$
acting between frequencies $\nu$, and $\nu'$.

We apply this formalism to study the sensitivity of 
parametrization of the 
%determination of the properties of 
cosmological anisotropy derived from the DMR 
data 
to the possible contamination by high latitude
galactic foreground emission.
We assume 
that 
%at high galactic latitude 
the faint foreground glow 
of synchrotron, free-free, and dust emission 
%in the Galaxy
outside the galactic plane
can be
traced by the available templates from observations at 
frequencies where the CMB does not dominate.
Two plausible  selections 
(Kogut \etal 1996a,b)
are the {\it DIRBE} 140 $\mu $m data (Reach \etal 1995), 
and the 408 MHz full sky radio map (Haslam \etal 1980).
(We also used the 110, and 240 $\mu $m {\it DIRBE} data and found the 
coupling of these to the DMR frequencies indistiguishable from that derived
at 140 $\mu $m.)
We conduct the analysis as outlined above
with $M=3$ DMR frequencies and $N=2$ templates. Together with
the CMB parameters $Q$ and $n$ this constitutes an 8-dimensional
likelihood problem. 
(A further extension of the method, described in Banday \etal 1996b,
addresses the issue of cross-correlation of the DMR data with existing
catalogues of extragalactic objects; there we have employed $N=6$ templates,
and solved the 20-dimensional likelihood problem.)
For each Galaxy cut selection we evaluate 
the $Q$-$n$ likelihood both without 
and with the foreground correction applied to the data. At each $Q$-$n$ 
grid point we 
derive the maximum likelihood solutions
$\widehat{\alpha}_\nu^i(Q,n)$ (and their covariance matrix), and 
use these values to evaluate the foreground corrected likelihood.
Maximization of the corrected likelihood renders 
simultaneously the most likely solution of both the cosmological
($Q$-$n$), and foreground ($\widehat{\alpha}_\nu^i$) fitting problem.
The likelihood surface in $Q$ and $n$ is sharply peaked
(as shown in G94b for 2-year data), and despite
it being non-Gaussian in functional form, the average values of the foreground
coupling constants
($\left\langle\widehat{\alpha}_\nu^i(Q,n)\right\rangle$, 
weighted by the corrected
likelihood) turn out practically identical to the maximum likelihood
solutions. The likelihood averaged values of  the six coupling
constants between 
the three DMR maps and the 140 $\mu$m and 408 MHz data
are shown in Fig. 2. 
The physical interpretation of these results in terms of the 
inferred contribution of galactic dust, free-free, and synchrotron
emission in the high-latitude DMR sky maps is discussed by Kogut
\etal (1996a,b). Here, we proceed to discuss the primary
target of the analysis of the foreground corrected DMR sky maps ---
the cosmic CMB anisotropy.

\section{RESULTS} 

Fig. 3 (Color plate) 
shows the coadded,
foreground corrected DMR sky maps
in both the ecliptic and galactic coordinate pixelization. 
For visualization the maps have been ``cleaned'' by removing 
the $\ell > 40$ noise (the $7^\circ$ FWHM DMR beam picks up sky signals 
that are entirely contained within $\ell \le 40$ in Fourier space). 
Ecliptic and galactic maps appear
very consistent at large angular scales, but near the pixel scale one can
notice the localised effects of different noise binning depending 
on the reference frame.
The choice of pixelization re-bins roughly half of the instrument noise
while leaving the CMB anisotropy unchanged;
consequently, slight variations in the likelihood results are expected.

Fig. 4 displays the cut sky DMR power spectra before and after the foreground 
correction evaluated for four combinations
of pixelization and Galaxy cut. (These spectra are shown here for illustration
purposes only; the likelihood method uses 
the Fourier coefficients $c_i$, not any quadratic function thereof.)
The apparently steep power spectrum
of the foreground Galaxy templates, $\sim \ell^{-3}$ at higher $\ell$, was 
discussed by Kogut \etal (1996a).  
An important result of the simultaneous CMB background $(Q,\,n)$ 
and foreground $(\alpha_\nu^k)$ fitting is
that the Galaxy correction
affects only the very low order multipoles of the derived cosmic CMB 
anisotropy.
Since the quadrupole moment of the Galaxy is partially counter-aligned to
the CMB quadrupole (Kogut et al.\ 1996b),
correcting the DMR maps for Galactic emission
{\it increases} the estimated CMB quadrupole.

Table 1 provides a comprehensive summary of the maximum likelihood solutions 
for the CMB anisotropy power spectrum model parameters.
The first entries in Table 1 refer to the high-latitude coadded DMR maps
from which no model of Galactic emission has been removed
and include, for comparison with the published 2-year analyses, the
results of re-application of the $\vert b\vert =20\ $deg Galaxy cut. 
One should note the following:
(1) compared to the G20 2-year results described in G94b, the 4-year
DMR data render a slightly smaller, but statistically consistent,
power sepctrum normalization (roughly a $\sim 0.5\; \mu$K drop of
$Q\vert_{n=1}$);
(2) solutions for $n$ are fairly stable with respect to
variation of the Galaxy cut, but the derived anisotropy amplitude 
decreases by more than 1 $\mu$K upon removal of the bright
Scorpius-Ophiucus and Taurus-Orion foreground regions (i.e.
upon extension of Galaxy cut from `20' to `20+'),
(3) an extra, but smaller,  decrese of the amplitude results from a further 
extension of Galaxy cut (to `30+'); we consider this as suggestive of 
having reached a sufficiently deep Galaxy plane excision with the `20+' cut.
Certainly the higher Galactic cut excludes real CMB features 
correlated over large  areas of the sky 
in the range $20^\circ < \vert b\vert < 30^\circ$, which
affects the $Q-n$ fits.

Comparing results for which the quadrupole is included ($\ell_{min}=2$)
to those from which it has been excluded ($\ell_{min}=3$),
we find again a systematic shift to higher normalization and lower index $n$
in the quadrupole-excluded results.  
The shift is well within the uncertainties from noise and cosmic variance,
but reflects at least in part the counter-alignment 
of the CMB and Galactic quadrupoles:
if the quadrupole is to be retained in the inference of cosmological 
parameters, Galactic emission should be accounted for in the data.

The next set of entries in Table 1 refers to the DMR maps
corrected for Galactic emission using 
the 408 MHz survey to trace synchrotron emission and the
DIRBE 140 $\mu$m map to trace dust emission.
Free-free emission is removed using a linear combination of the
dust- and synchrotron- corrected DMR maps.
This map makes no assumptions as to the distribution of free-free emission,
but is relatively noisy compared to the coadded maps, which is reflected in
lower significance of all the derived parameters.
The final set of entries refers to the maximally sensitive coadded maps,
corrected for Galactic emission using the 408 MHz survey to trace synchrotron 
emission and the DIRBE 140 $\mu$m map to trace both dust and free-free emission.
The fitted values for $Q$ and $n$ are in good agreement for both 
foreground correction techniques.
One will note that the high latitude foreground correction 
further reduces the fitted power 
spectrum amplitude by $\sim 0.5\ \mu$K.
This cannot be considered as significant particularly in the light of a
similar order of discrepancy between the galactic and ecliptic map
maximum likelihood solutions, and especially in comparison with an overall
statistical significance of all the determined parameters.

\section{SUMMARY}

A statistically unbiased technique of 
power spectrum determination 
%using cut-sky Fourier decomposition of the high-latitude portion of the sky
has been used to infer the CMB normalization 
$Q_{rms-PS}$ and power-law index $n$
from the {\it COBE} DMR 4-year sky maps.
The sky maps are decomposed 
in the basis of orthonormal functions on the cut sky
into a set of coefficients linear in the pixel temperatures.
These coefficients are then used in a
likelihood analysis to infer the parameters of the 
CMB anisotropy models.
We account for high-latitude Galactic emission by
using the 408 MHz sky survey to trace the angular distribution
of synchrotron emission and the DIRBE 140 $\mu$m map to trace dust emission.
%and fitting for the amplitude of emission in the DMR maps traced by these 
%templates.
Free-free emission is removed using either a linear combination of the DMR maps 
or using the DIRBE map to remove free-free as well as dust emission.
Since we find some minor changes in the fitted parameters 
with the Galaxy cut or map pixelization and method of high latitude
foreground modeling, 
a judgement call is required to quote a unique {\it COBE} DMR
determination of the primordial power spectrum parameters.
The 4-year data are consistent with
quadrupole normalization $Q_{rms-PS} \sim 15.3^{+3.7}_{-2.8} ~\mu$K
and power-law index $n \sim 1.2 \pm 0.3$.
With $n$ fixed at unity
we derive the normalization
for an exact scale-invariant Harrison-Zel'dovich power spectrum
$Q_{rms-PS}|_{n=1} \sim 18 ~\mu$K with high ($\simgt 14\sigma$)
significance $\delta Q/Q \simlt 0.07$.
%(all errors are quoted conservatively).
Implications of the {\it COBE} DMR 4-year data for the specific models of
large scale structure formation will be discussed elsewhere.

\ms
%\noi{ACKNOWLEDGEMENTS}

We gratefully acknowledge the efforts of those contributing to the {\it COBE}
DMR. {\it COBE} is supported by the Office of Space Sciences of NASA
Headquarters.

\clearpage
\begin{center}
FIGURE CAPTIONS
\end{center}

\vspace{7mm}
\noindent Fig. 1 (Color Plate). 
An example of the full-sky spherical harmonic and the related 
cut-sky orthonormal functions (G\'orski 1994)
--- note the coordinate frame and Galaxy cut dependence ---
used in Fourier analysis of the DMR and foreground emission template data.

\vspace{7mm}
\noindent Fig. 2. Coefficients of coupling between the three DMR channels
at 31.5, 53, and 90 GHz and the 
Galactic templates: the 
140$\mu $m {\it DIRBE} data --- filled symbols, and the 408 MHz radio data ---
open symbols.
The coefficients shown in the plot (computed for the choice of 
pixelization and Galaxy cut as indicated)
are the maximum likelihood
values averaged over the $(Q-n)$ likelihood of the 
%$n$ and $Q$
%$Q_{RMS-PS}$ 
simultaneous background and foreground
fit to the 4-year DMR data.
The one $\sigma$ confidence intervals are defined
in the  usual manner as the square roots of the diagonal on the 
covariance matrix.

\vspace{7mm}
\noindent Fig. 3 (Color Plate). {\it COBE} DMR 4-year,
31.5, 53, and 90 GHz  sky maps coadded after 
foreground correction according to the 140$\mu $m- and 408 MHz-template 
fitting.
Galactic and ecliptic 
(rotated to galactic coordinates to facilitate the comparison)
pixelizations are shown. Noise reducing Fourier removal of the $\ell > 40$ 
content of the raw data
has been 
applied to the full sky maps in order to aid visualization. 
Apparent small differences in pixel temperatures are induced by 
the reference frame dependent time ordered data noise binning in the map 
making procedure.

%\clearpage
%\begin{figure}
%\plotone{ff1.eps}
%\caption{ }
%\end{figure}

\vspace{7mm}
\noindent Fig. 4. Power spectra of the coadded DMR data and the templates
used to assess the galactic foreground emission. 
The four panels exhibit  the cut-sky rms power spectra,
%$\biggl(
$c_\ell \equiv 
\Bigl[ \left( \sum_{i=\ell^2 +1}^{i=(\ell+1)^2} c_i^2\right) / 
(2\ell +1)\Bigr]^{1/2}$,
%\biggr)$ 
evaluated for both ecliptic and galactic data with
two Galaxy cuts as indicated. (Small letter $c_\ell$ is used in the definition
of the cut-sky spectrum
to avoid confusion with commonly used $C_\ell$ symbols, which denote the
theoreticall, full-sky anisotropy spectrum.)
Open circles --- coadded
raw DMR data, filled circles --- template corrected DMR data, filled squares 
--- foreground emission contribution to the DMR 53 GHz data characterized by 
spatial structure of the 140$\mu $m DIRBE map,
arrows --- $95\%$ upper limit to possible foreground emission in the DMR 53 GHz
channel characterized by spatial structure of the 408 MHz radio data,
horizontal lines --- rms noise per Fourier mode in the coadded 4-year DMR
maps.


\clearpage
\begin{references}


\rep Banday, A.J., et al.;1996a;ApJ (submitted)

\rep Banday, A.J., et al.;1996b;ApJ (submitted)

\ref Bennett, C.L., et al.;1992;ApJ;396;L7

\rep Bennett, C.L., Banday, A.J., G\'orski, K. M., Hinshaw, G.,
Kogut, A., \& Wright, E.L.;1996;ApJ (submitted)

\ref Bond, J. R., \& Efstathiou, G.;1987;MNRAS;226;655

\ref de Vaucouleurs, G.;1956;Vistas in Astronomy;2;1584

\ref Fabbri, R.,  Lucchin, F., \& Matarrese, S.;1987;ApJ;315;1

\ref G\'orski, K.M.;1994;ApJ;430;L85~(G94a)

\ref G\'orski, K.M., et al.;1994;ApJ;430;L89~(G94b)

\rep G\'orski, K.M., et al.;1996;ApJ (in preparation)

\ref Haslam, C.G.T, Klein, U., Salter, C.J., Stoffel, H, Wilson, W.E., Cleary, 
M.N., Cooke, D.J., \& Thomasson, P.;1981;A\&A;100;209

\ref Kogut, A., et al.;1992;ApJ;401;1

\rep Kogut, A., Banday, A.J., Bennett, C.L., G\'orski, K. M.,
Hinshaw, G., \& Reach, W.T.;1996a;ApJ (in press)

\rep Kogut, A., Hinshaw, G., Banday, A.J., Bennett, C.L., 
G\'orski, K. M., \& Smoot, G.F.;1996b;ApJ (submitted)

\rep Kogut, A., et al.;1996c;ApJ (submitted)

\rep Kogut, A., Banday, A.J., Bennett, C.L., 
G\'orski, K. M., \& Hinshaw, G.;1996d;ApJ (submitted)

\ref Lineweaver, C., et al.;1994;ApJ;436;452

\book Reach, W.T., Franz, B.A., Kelsall, T., \& Weiland, J.L.;1995;
Unveiling the Cosmic Infrared Background, ed.\ E.\ Dwek, (New York:AIP)

\ref Smoot, G. F. et al.;1992;ApJ;396;L1

\ref Wright, E.L., et al.;1992;ApJ;396;L13

%\rep Wright, E.L., et al.;1996;ApJ (submitted)

\end{references}
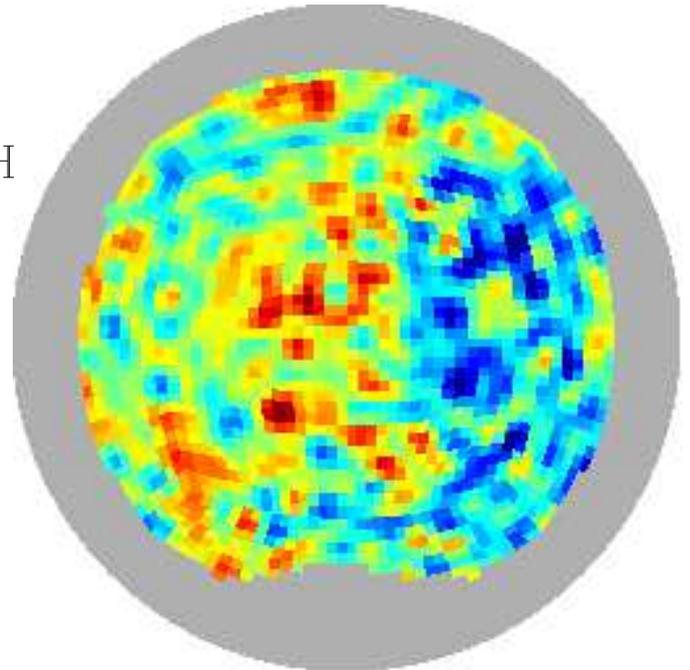
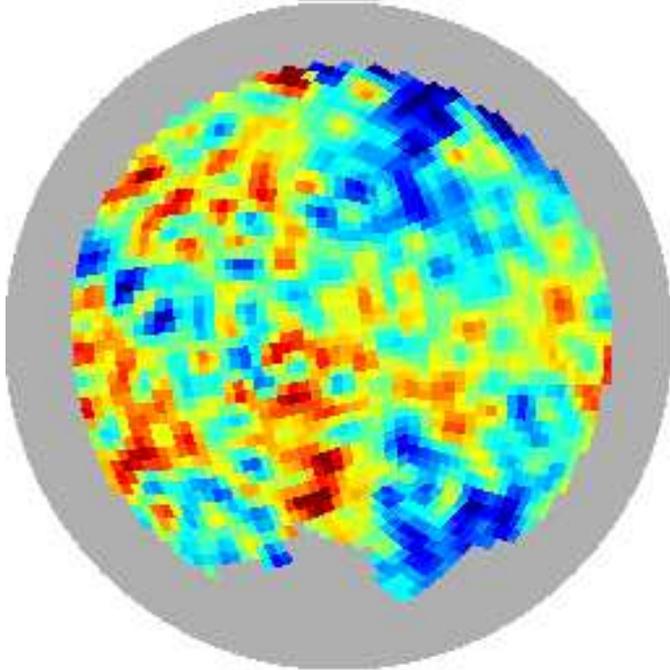
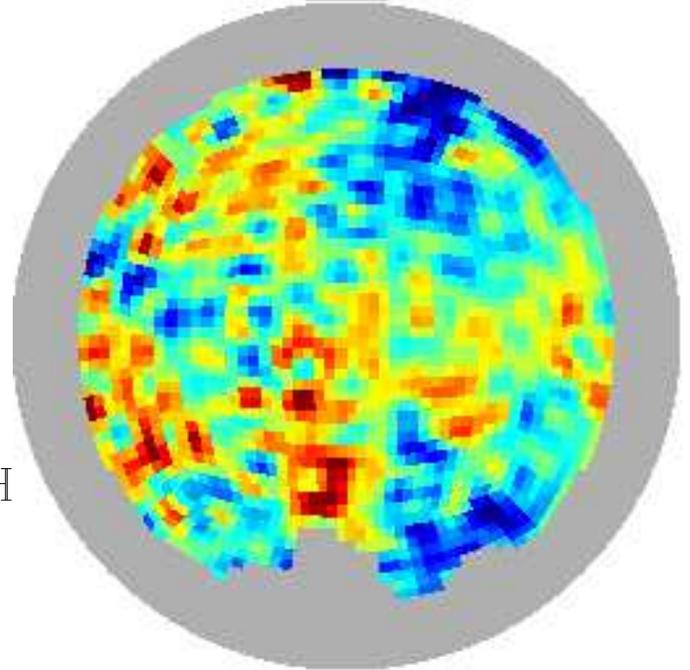
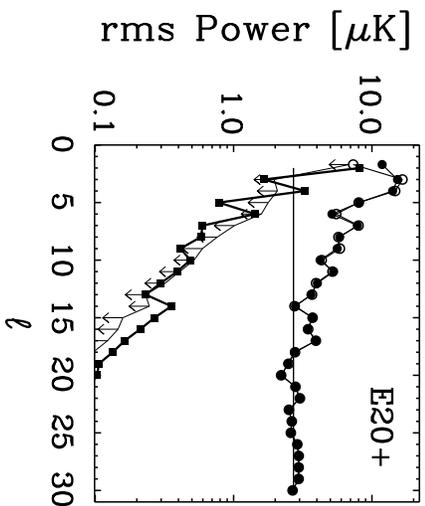
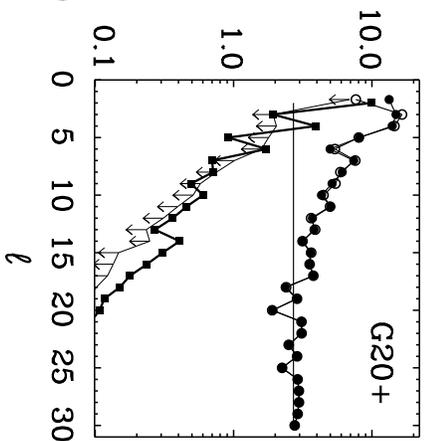
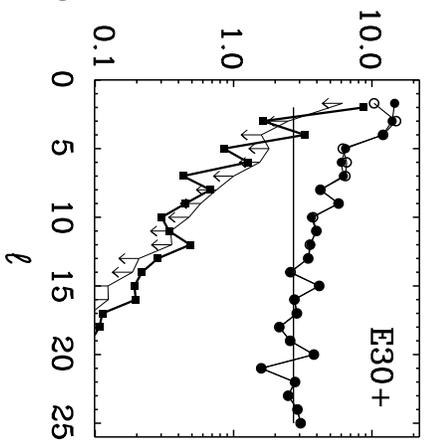
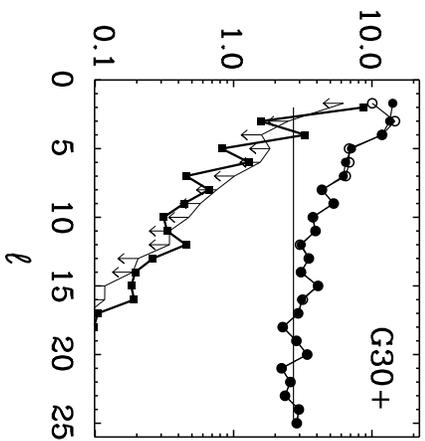
\end{document}